\documentclass[11pt]{article}
\usepackage{setspace,braket}
\usepackage{amsmath, amssymb, bm, yfonts}
\usepackage{etoolbox}
\usepackage{jheppub}
    \makeatletter
    \patchcmd{\maketitle}{\@fpheader}{}{}{}
\makeatother

\def\be{\begin{equation}}
\def\ee{\end{equation}}

\def\bg{\bar{g}}
\def\beq{\begin{eqnarray}}\def\eeq{\end{eqnarray}}
\def\ba#1\ea{\begin{align}#1\end{align}}
\def\bg#1\eg{\begin{gather}#1\end{gather}}
\def\bm#1\em{\begin{multline}#1\end{multline}}
\def\bmd#1\emd{\begin{multlined}#1\end{multlined}}

\def\({\left(}
\def\){\right)}
\def\[{\left[}
\def\]{\right]}

\usepackage{float} 
\usepackage{amsmath}
\usepackage{amsfonts}   
\usepackage{amssymb}

\begin{document}
\title{ Viscosity bound for anisotropic superfluids in higher derivative gravity}
\author[a]{Arpan Bhattacharyya }
\author[a]{and Dibakar Roychowdhury}
\affiliation[a]{Centre for High Energy Physics, Indian Institute of Science, C.V. Raman Avenue, Bangalore 560012, India}
\emailAdd{arpan@cts.iisc.ernet.in,dibakarphys@gmail.com, dibakar@cts.iisc.ernet.in}

\begin{abstract}
{In the present paper, based on the principles of gauge/gravity duality we analytically compute the shear viscosity to entropy ($ \eta/s $) ratio corresponding to the super fluid phase in Einstein Gauss-Bonnet gravity. From our analysis we note that the ratio indeed receives a finite temperature correction below certain critical temperature ($ T<T_c $). This proves the non universality of  $ \eta/s $ ratio in higher derivative theories of gravity. We also compute the upper bound for the Gauss-Bonnet coupling ($ \lambda $) corresponding to the symmetry broken phase and note that the upper bound on the coupling does not seem to change as long as we are close to the critical point of the phase diagram. However the corresponding lower bound of the $ \eta/s $
ratio seems to get modified due to the finite temperature effects.}
 \end{abstract}
\maketitle
\onehalfspace
\section{Overview and Motivation}

In the recent years the AdS/CFT correspondence \cite{jm} has been found to provide an excellent framework to study various properties of a large class of strongly coupled gauge theories that admit a dual description in the form of the classical theory of gravity defined in asymptotically Anti-de-Sitter (AdS) space time. One of the significant achievements of AdS/CFT duality is that it provides a universal lower bound for the shear viscosity to entropy ratio namely,
\begin{eqnarray}
\frac{\eta}{s} =\frac{1}{4\pi}
\end{eqnarray}
for a wide class of strongly coupled field theories that could be described in terms of the usual two derivative Einstein gravity as the dual counterpart \cite{Son3,Alex1,Alex2, Liu1}.
This universal bound has remarkable agreement to that with the experimentally measured shear viscosity in various systems like the quark gluon plasma or the cold atom systems at unitarity \cite{shuryak, talks, rick}.

As per as the real world systems are concerned, it has been known for quite some time that the shear viscosity to entropy ratio is not universal in the sense that it depends on the temperature of the system. This observation was in fact enough motivating to look forward for certain theoretical framework that eventually explains this crucial experimental fact.  At this stage it is worthwhile to mention that one such (theoretical) attempt to address the above observational fact comes from the standard prescription of gauge/gravity duality. 

Recently in certain holographic calculations it has been observed that for $ p $-wave (non abelian) superfluids the above universality of the $ \eta/s $ ratio seems to get violated below certain critical temperature ($ T<T_c $) \cite{Basu1, Johanna1, Natsuume,Basu2}. This is due to the fact that in the presence of the $ p $- wave order parameter the spatial rotational symmetry times the $ U(1) $ rotational symmetry generated by one of the $ SU(2) $ generator namely $ \tau^{3} $ gets spontaneously broken in the bulk. As a result the global $ SO(3) $ symmetry of the boundary theory gets spontaneously broken to the global $ SO(2) $ symmetry which therefore results in a spatial anisotropy for the boundary theory \cite{Manvelyan,Gubser,Roberts,Herzog1,Bhattacharya}. 

Motivated from the above analysis, the purpose of the present article is to compute the $ \eta/s $ ratio corresponding to the symmetry broken phase considering the Einstein Gauss-Bonnet (EGB) gravity coupled to Yang-Mills field in $ AdS_{5} $. Computation of $ \eta/s $ ratio in higher derivative theories of gravity is an interesting project in itself since the universality of the lower bound does get violated automatically without going into any super-fluid/symmetry broken phase of the system \cite{kp,blmsy, Alex3, Myers2, Nabankar,rami,cs,others,Crem}\footnote{There exists a huge literature on this topic. For more interesting aspects of bounds of $\eta/s$ ratio interested readers are referred further to  \cite{chls,ghodsi,junk,vi1,David,erd,Buchel}.}. The question that we would mostly like to  address in this article could be stated as what is the corresponding finite temperature corrections to shear viscosity to entropy ($\eta/s$) ratio in EGB gravity. In other words, the kind of question that we are going to address in this paper eventually raises the fact that the result for the $ \eta/s $ ratio in the context of EGB gravity should also receive some finite temperature corrections like in the case for the  usual  two derivative theory of gravity  \cite{Basu1, Johanna1, Natsuume,Basu2}.

In our analysis we consider the symmetry to be broken explicitly in the $ (x,y) $ plane while it preserves  symmetry in the $ (y, z) $ plane. Therefore a natural expectation would be to find the corresponding $ \eta_{yz}/s $ ratio to be unchanged and it should match with the earlier observations in EGB gravity \cite{blmsy}. Whereas, on the other hand, some interesting physics should emerge while we compute the $ \eta_{xy}/s $ ratio corresponding to the symmetry broken phase since the interaction between the gravitons to that with the gauge bosons in the symmetry broken phase should yield some finite temperature corrections to the existing result for $ \eta/s $ ratio in EGB gravity. In other words, the ratio should get modified due to the fact that the shear modes corresponding to the $ (x,y) $ plane are no more helicity two excitations rather they transform like helicity one or vector modes of the unbroken $ SO(2) $ symmetry group which interact with other helicity one modes of the theory namely the gauge bosons in the symmetry broken phase.

The second motivation of our analysis comes from the fact that in EGB theory of gravity the GB coupling ($ \lambda $) is actually constrained by the fact that the theory has to satisfy the causality namely the velocity of the graviton wave packet cannot exceed the velocity of photon \cite{blmsy}\footnote{ For more aspects of causality constraints interested readers are referred to  \cite{hm}.}. This eventually motivates us to explore the bound on $ \lambda $ corresponding to the unbroken as well as the symmetry broken phase in the presence of Yang-Mills coupling. Finally, it should be noted that throughout the analysis we treat the Gauss-Bonnet (GB) coupling ($ \lambda $) non perturbatively while on the other hand we treat the $ SU(2) $ gauge sector perturbatively.

The organization of the paper is the following: In Section 2 we discuss the basic holographic set up where we consider the Einstein Gauss-Bonnet (EGB) gravity coupled to Yang-Mills field in the presence of back reaction. In Section 3 we compute the $ \eta/s $ ratio corresponding to the symmetry broken phase. In Section 4 we explore the causality constraint on GB coupling ($ \lambda $) corresponding to the symmetry broken as well as symmetry unbroken phase. Finally we conclude in Section 5.

\section{The holographic set up}

Before we actually start our analysis, we would first like to note down all the crucial assumptions as well as the approximations that have been taken care of in the subsequent computations. The purpose of the present section is to provide a detail discussion on the holographic set up as well as the solutions of the field equations in the bulk. This might be regarded as the first step towards constructing the anisotropic superfluid phase at the boundary. 

In our analysis we consider the Gauss-Bonnet (GB) coupling ($ \lambda $) to be non perturbative whereas on the other hand, we treat the Yang-Mills sector to be perturbative near the critical point ($ T \sim T_c $). This construction has two notable features - Firstly, it provides us with the exact framework to compare the $ \eta/s $ ratio corresponding to the anisotropic/symmetry broken phase to that with the earlier results known for the symmetry unbroken phase \cite{Basu2}. Secondly with this assumption one can easily construct solutions to the field equations near the UV as well as the IR sector of the theory. This is in fact sufficient for the present analysis since the only entity that we finally want to compute is the retarded Green's function for the boundary theory.

We start our analysis by considering Einstein Gauss-Bonnet (EGB) gravity coupled to $SU(2)$ Yang-Mills field in an asymptotically $AdS_5$ space time. The corresponding action reads as,
\be
S=\int d^{5}x\sqrt{-g} \,\,\left[ \frac{1}{2\kappa_{5}^2}\,\left [ R+ \frac{12}{L^2} + \frac{\lambda L^2}{2} \Big(R_{\alpha\beta\gamma\delta}R^{\alpha\beta\gamma\delta}-4 R_{\alpha\beta}R^{\alpha\beta}+ R^2 \Big)\right]  -\frac{1}{4 g^2} F_{\alpha \beta}^{a} F^{a\alpha\beta}\right] \\
 \ee
where the field strength tensor could be formally expressed as,
\be
F^{a}_{\alpha\beta}=\partial_{\alpha} A^{a}_{\beta}-\partial_{\beta} A^{a}_{\alpha}-\epsilon^{abc}A^{b}_{\alpha} A^{c}_{\beta}\,.\\
\ee

The corresponding Einstein and Yang-Mills equations of motion turn out to be,
\begin{eqnarray}
R_{\alpha \beta }+ 4 g_{\alpha \beta}-\frac{\lambda}{6} \Big [g_{\alpha\beta}\Big( R_{\gamma\delta\mu\nu}R^{\gamma\delta\mu\nu}-4 R_{\mu\nu}R^{\mu\nu}+ R^2 \Big)+H_{\alpha\beta} \Big]-\kappa_{5}^2\Big (T_{\alpha\beta}-\frac{1}{3} T g_{\alpha\beta}\Big)&=&0\nonumber\\
\nabla_{\beta} F^{a \beta}_{\alpha}-\epsilon^{abc} A^{b}_{\beta} F^{\beta}_{\alpha}&=&0\nonumber\\
\label{E4}
\end{eqnarray}
respectively\footnote{We will set L=1 for convenience.}. The function $H_{\alpha\beta}$ as well as the energy momentum tensor ($T_{\alpha\beta}$) take the following form namely,
\begin{eqnarray}
H_{\alpha\beta}&=& 2 R_\alpha^\delta  R_{\beta \delta}-2 R^{\delta \sigma} R_{\delta \alpha \beta \sigma}-  R  R_{\alpha\beta} - R_{\alpha \sigma \delta \mu}{ R_\beta}{}^{\sigma \delta 
  \mu}\nonumber\\
  T_{\alpha\beta}&=& \frac{1}{g^2}\Big( F^{a}_{\alpha\gamma}F^{a \gamma}_{\beta}-\frac{1}{4}F_{\mu\nu a}F^{\mu\nu a} g_{\alpha\beta}\Big). 
\end{eqnarray}

The metric ansatz as well as the ansatz for the $ SU(2) $ gauge field for our analysis could be formally expressed as,
\begin{eqnarray}
ds^2&=&  -N(r) \sigma(r)^2 N_{\#}^2 dt^2+\frac{1}{N(r)} dr^2+ \frac{r^2}{L^2} f(r)^{-4} dx^2 + \frac{r^2}{L^2} f(r)^2( dy^2+ dz^2)\nonumber\\
A&=&\phi(r) \tau^3 dt^2+ \omega(r) \tau^1 dx
\end{eqnarray}
where $ \omega(r) $ corresponds to the $ p $- wave order parameter that spontaneously breaks the $ SO(3) $ symmetry of the boundary theory below $ T<T_c $. Here $ N_{\#}^2$ is a an arbitrary constant. In our analysis we set  \be  N_{\#}^2=\frac{1}{2}(1+\sqrt{1-4\lambda})\ee so that the velocity of light at the boundary of the $ AdS_5 $ becomes unity \cite{blmsy}. Also we could define the ratio of the AdS length scale to that with the cosmological length scale as\footnote{ Here we closely follow the notation of \cite{Myers:2010ru}. For a detailed discussion on $f_{\infty}$ interested readers are referred to \cite{Myers:2010ru}, in particular the equations 2.7 and 2.8 of the first of the references provided in \cite{Myers:2010ru}.}, 
\be
f_{\infty}=\Big( \frac{L_{AdS}}{L}\Big)^2\\
\ee 
which satisfies the following equation of motion,
\be
1-f_{\infty}+ f_{\infty}^2\lambda=0\,.\label{E10}
\ee
Eq.(\ref{E10}) will in general have two roots of which we take the one which smoothly connects with the Einstein case ($\lambda=0$) when $f_{\infty}$ goes to one. 

Our next goal would be to solve (\ref{E4}) in the large $r$ limit. In order to do that we first consider the following perturbative expansion of the gauge field namely,
\begin{eqnarray}
A_{\mu}^{a}= A_{\mu}^{a(0)}+\varepsilon  A_{\mu}^{a(1)}+\varepsilon^{2}  A_{\mu}^{a(2)}+O(\varepsilon^{3})\label{E11}
\end{eqnarray}
where $ \varepsilon(=(1-T/T_c)) $ is a small \textit{positive} dimensionless parameter such that $ \mid \varepsilon \mid << 1 $. The expansion (\ref{E11}) in fact reflects the fact that we are essentially considering our system (superfluid) at a temperature $ T $ which is very close to the critical temperature ($ T\sim T_c $). In the next step we shall further expand each of the terms on the r.h.s of (\ref{E11}) as a perturbation in the parameter $ \alpha^{2}(=\frac{\kappa_{5}^2}{4 g^2}) $ namely,
\begin{eqnarray}
A_{\mu}^{a(i)}=\mathcal{A}_{\mu}^{a(i)(\alpha^{(0)})}+\alpha^{2}\mathcal{A}_{\mu}^{a(i)(\alpha^{(2)})}+O(\alpha^{4}).\label{alpha}\label{E12}
\end{eqnarray}

Using the above expansions (\ref{E11}) and (\ref{E12}) we finally solve Eq.(\ref{E4}) separately near the UV as well as the IR sector of the theory as it is in fact quite difficult to solve Eq.(\ref{E4}) for any generic value of the radial coordinate ($r$). The solutions corresponding to the large values of the radial coordinate ($ r $) turn out to be,
\begin{eqnarray}
\sigma(r)&=&1-\varepsilon^2\alpha^2\frac{2}{9r^6}\, {\ \ }f(r)=1+\varepsilon^2\alpha^2\frac{1}{9r^6}\nonumber\\
N(r)&=&f_{\infty}\, r^2-\frac{1}{\sqrt{1-4\, \lambda } \,r^2} +\mathcal{O}\Big(\frac{1}{r^6}\Big)\nonumber\\&+&\frac{32\alpha^2}{3}\Big(-\frac{1}{\sqrt{1-4 \,\lambda } \,N_{\#}^2 \,r^2}+\frac{1}{ \sqrt{1-4 \, \lambda }\, N_{\#}^2  r^4}+\mathcal{O}\Big(\frac{1}{r^6}\Big)\Big)\nonumber\\&-&\varepsilon^2\alpha^2 \Big(\frac{281}{1260 r^2}+\frac{ f_{\infty }\,\left(840 N_{\#}^2+281\right)}{1260 N_{\#}^2 \left(f_{\infty }-2\right)}+\mathcal{O}\Big(\frac{1}{r^6}\Big)\Big)\nonumber\\
\omega(r)&=&\varepsilon\Big(\frac{1}{r^2}-\frac{2\,(  N_{\#}^2-\lambda )}{N_{\#}^2\ r^4}+\mathcal{O}\Big(\frac{1}{r^6}\Big)\Big)+ O(\varepsilon^2)\nonumber\\ 
\phi(r)&=&4\Big(1-\frac{1}{r^2}\Big)+\varepsilon^2 \left(\frac{71}{6720}-\frac{281}{6720\,\, r^2}+\mathcal{O}\Big(\frac{1}{r^6}\Big)\right).\label{Sol}
\end{eqnarray}

A number of comments are to made at this stage. Firstly, for the computation of $\eta/s$ we only need the metric upto $\mathcal{O}(\frac{1}{r^4}).$  For that it is sufficient to expand $f(r)$ upto $\mathcal{O}(\frac{1}{r^6})$ as in the metric components it comes with a $r^2$ factor multiplying it. We have expanded $N(r)$ upto $\mathcal{O}(\frac{1}{r^4})$. Now as the leading term of $N(r)$ goes as $r^2$ we have to expand $\sigma(r)$ upto $\mathcal{O}(\frac{1}{r^6})$ as in the metric they come in the form of  $N(r)\sigma(r)^2\,.$  Secondly we have set the radius of the horizon equal to unity namely, $r_{h}=1$ and lastly the solutions mentioned above in (\ref{Sol}) are exact in the Gauss-Bonnet (GB) coupling ($ \lambda $).  

Next we perform the near horizon expansion in order to obtain solutions corresponding to the IR sector of the theory. The solutions thus obtained are found to be valid upto $\mathcal{O}(r-1)$. In the following we quote the near horizon solutions that turn out to be\footnote{We require the near horizon data in order to compute the temperature ($ T $) of the black brane correctly. Near horizon solutions also play an important role to understand the near horizon behaviour of the graviton as well as the gauge fluctuations to compute the shear viscosity to entropy ratio.}, 
\begin{eqnarray}
\sigma(r)&=&1+ \varepsilon^2\alpha^2 \Big (-\frac{1}{36}+\frac{r-1}{12}\Big);~~f(r)=1+\varepsilon^2\alpha^2\Big (\frac{1}{288}\Big);~~ \omega(r)=\frac{\varepsilon}{4}\nonumber\\
N(r)&=&4 (r - 1) -\frac{64\alpha^2}{3\,N_{\#}^2} ( r-1)
+\varepsilon^2\alpha^2 \frac{52 (208+315 N_{\#}^2)}{54915\,N_{\#}^2 (3-16 \,\lambda )}(r-1)\nonumber\\
\phi(r)&=&8 (r - 1)-\frac{13\, \varepsilon^2 }{420} (r-1).\label{Sol1}
\end{eqnarray}
The temperature of the black brane is defined as usual by defining the surface gravity.  The temperature is defined as usual by the  formula shown below. 
\be  \label{T}
T= \frac{1}{2\pi}\sqrt{(-\frac{1}{4} g^{tt}g^{rr}(\partial_{r} g_{tt})^{2})}\,\,\Bigr \rvert_{ r=r_{h}=1}
\ee
Using (\ref{Sol1}) in (\ref{T}) and expanding upto $\mathcal{O}(\varepsilon^2\alpha^2)$ we get the black brane temperature as, 
\be \label{T1}
T=\frac{ N_{\#}}{\pi }\Big( 1-\frac{16\, \alpha^2}{3  N_{\#}^2}-\frac{4 \varepsilon^2 \alpha^2}{9(3-16\lambda)} \Big( \frac{165 }{8368}-\frac{2028}{18305 N_{\#}^2}- \lambda \Big)\Big).  
\ee
The critical temperature at which the system undergoes a second order phase transition turns out to be\footnote{At this point one should note that for $ f_{\infty}=1 $ and $N_{\#}=1$, the expression for T and  $ T_c $ match exactly with the corresponding expressions of the two derivative Einstein gravity \cite{Basu2}.},
\begin{eqnarray}
T_c = \frac{N_{\#}}{\pi }\Big( 1-\frac{16\, \alpha^2}{3 N_{\#}^2} \Big).
\label{E215}
\end{eqnarray}
 Note that in order to obtain the critical temperature (\ref{E215}) what one essentially needs to do is to turn off the order parameter ($ \omega =0 $) which essentially instructs us to set $\varepsilon=0$ in (\ref{T1}) (see \ref{Sol}). The reason for this is the fact that the critical temperature marks the phase transition point where we have no symmetry breaking parameter as we have an exact  scale symmetry there. So turning off the symmetry breaking parameter i.e the $\varepsilon$ in (\ref{T1}) we identify the the critical temperature $T_{c}$. Also notice that $T_{c}$ explicitly contains the information about the gauge coupling constant inside the parameter $ \alpha^{2}(=\frac{\kappa_{5}^2}{4 g^2}) $ which has been defined as the ratio of the Newton's constant to that of the Yang-Mills couplings.
 
Finally, the entropy of the black brane could be formally expressed as \cite{Cai}, 
\begin{equation}
\label{Anisotrophic entropy}
S=\frac{2\pi}{\kappa^2_5}V_{3}
\end{equation}
where $V_{3}(=\int dxdydz)$ is the world volume of the black brane. 
With the above machinery in hand, we are finally in a position to compute the shear viscosity to entropy ($ \eta/s $) ratio corresponding to both the symmetry broken as well as the symmetry unbroken phase in presence of higher derivative corrections. This is basically the goal of our next section.

\section{ $ \eta/s$ for anisotropic superfluid }
In this section, using the Kubo's formula, we compute the shear viscosity to entropy ratio for anisotropic superfluids in Einstein Gauss-Bonnet (EGB) gravity. In order to do that we turn on fluctuations of the metric as well as the gauge fields namely $ h_{\mu\nu} $ and $ \delta A^{a}_{\mu}\,. $  The gravity fluctuations like $ h_{yz} $ and $ h_{yy}-h_{zz} $  transform as the tensor modes of unbroken $ SO(2) $ symmetry group. On the other hand the fluctuations like $ h_{xy}\,,\delta A^{1}_{y}$ and $\delta A^{2}_{y} $ transform as the vector modes of $SO(2)$. This basically suggests the fact that the graviton fluctuations along the $(x, y)$ plane do not transform as pure helicity two states of $SO(2)$ rather they are coupled with the gauge fluctuations which eventually results in some finite temperature corrections to shear viscosity bound.

\subsection{Calculation of $ \eta_{yz}/s$}

Let us first consider the graviton fluctuations along ($ y,z $) plane namely $h_{yz}$. 
We start with the linearized equation of motion and set,  \be h_{yz}(r) \equiv r^2 f^2(r)\Phi(r,t).\ee

Using Fourier transformation we can write down the field $\Phi(r,t)$ as,
\begin{equation}
\Phi(r,t)=\int^{\infty}_{-\infty}e^{-i\nu t}\Phi_{\nu}(r) d\nu.\label{E16}
\end{equation}

Substituting (\ref{E16}) in to (\ref{E4}) one finally arrives at some second order differential equation of $ \Phi_{\nu}(r)$ which could be read off as,
\begin{eqnarray}
0&=&\Big (\frac{\lambda \, N'(r)}{r}-1\Big)\Phi^{\prime\prime}_{\nu}(r)+\Big (\frac{\lambda\, N''(r)}{r}+\frac{2 \lambda\, N'(r)}{r^2}+\frac{\lambda\,  N'(r)^2}{r N(r)}-\frac{N'(r)}{N (r)}-\frac{3}{r} \Big)
\Phi^{\prime}_{\nu}(r)\nonumber\\&+&\Big(\frac{4 \lambda\, N'(r)}{r^3}+\frac{8}{N(r)}-\frac{2 N'(r)}{r N(r)}-\frac{2 \alpha ^2  \phi '(r)^2}{3 N_{\#}^2 N(r) \sigma (r)^2}-\frac{4}{r^2}\Big)\Phi_{\nu}(r)\nonumber\\ &+& \nu^2 \Big (\frac{\lambda\,   N'(r)}{r N_{\#}^2 N(r)^2 \sigma (r)^2}-\frac{1}{N_{\#}^2 N(r)^2 \sigma (r)^2}\Big) \Phi_{\nu}(r) \label{E18}
\end{eqnarray}
where the prime denotes derivative in the radial coordinate ($ r $).

In order to solve the full equation of motion corresponding to the graviton perturbation $ h_{yz} $, we consider an ansatz for $ \Phi_{\nu}(r) $ that basically reflects the incoming wave boundary condition for graviton modes namely,
\be
\Phi_{\nu}(r)= \left( \frac{N(r)}{r^2} \right)^{-i\frac{\nu \tilde T}{4}} F(r)\label{PHI}
\ee
where \be \tilde  T=\frac{1}{ N_{\#}}\Big( 1+\frac{16\, \alpha^2}{3  N_{\#}^2}+\frac{4 \varepsilon^2 \alpha^2}{9(3-16\lambda)} \Big( \frac{165 }{8368}-\frac{2028}{18305 N_{\#}^2}- \lambda \Big)\Big). \label{tt} \ee
 We then proceed to calculate the retarded Green's function using the recipe of \cite{Min}.  
In order to do that, as a first step we substitute (\ref{PHI}) in to (\ref{E18}) which yields,
\begin{eqnarray}
0&=&\Big(\lambda N'(r)-r\Big)F''(r)+\Big(1-\frac{r N'(r)}{N(r)}-\frac{2\,\lambda N'(r)}{r}+\frac{\lambda\, N'(r)^2}{N(r)}+\lambda N''(r)\Big)F'(r)\nonumber\\&+&\frac{i \nu \tilde T}{2}\Big(\frac{r N'(r)}{N(r)}-2+\frac{2 \,\lambda N'(r)}{r}-\frac{\lambda\, N'(r)^2}{N(r)}\Big)F'(r)+\Big(\frac{8 r}{N(r)}-\frac{4}{r}+\frac{6\, \lambda\, N'(r)}{r^2}- \nonumber\\&& \frac{2\lambda N'(r)^2}{r N(r)}-\frac{2\,\lambda\, N''(r)}{r}\Big)F(r)-\frac{128\,\alpha^2N(r)+ 3 r^5\nu^2(r-\lambda N'(r))}{3 r^5N(r)^2\sigma(r)^2N_{\#}^2} F(r)\nonumber\\&+&
\frac{i\nu \tilde T}{4}\Big(\frac{4}{r}-\frac{3N'(r)}{N(r)}+\frac{r N''(r)}{N(r)}-\frac{6\,\lambda\,N'(r)}{r^2}+\frac{4\,\lambda\,N'(r)^2}{r N(r)}+\frac{2\,\lambda\,N''(r)}{r}-\nonumber\\&&\frac{2\,\lambda\,N'(r)N''(r)}{N(r)}\Big)F(r)-\frac{\nu^2 \tilde T^2}{16}\Big(\frac{4 N'(r)}{N(r)}-\frac{4}{r}-\frac{r N'(r)^2}{r N(r)}+\frac{4\lambda N'(r)}{r^2}-\nonumber\\&& \frac{4\,\lambda\,N'(r)^2}{r N(r)}+\frac{\lambda\,N'(r)^3}{N(r)^2}\,. \label{eq21}
\end{eqnarray}

In order to solve the above equation (\ref{eq21}), as a first step we expand $F(r) $ perturbatively in $\nu$ as,
\be
F(r)= F_{0} (r)+\frac{i \nu}{4} F_{1}(r)+\mathcal{O}(\nu^2).
\ee
In the next step we expand each of the $F_{0}(r)$ as well as $ F_{1}(r)$ around the boundary and we get,
\be
F(r)= \Phi_{0}+ \frac{\Phi_{2}}{r^4}+ \mathcal{O}\Big(\frac{1}{r^6}\Big)+\frac{i \nu (1-4\,\lambda)}{4 N_{\#}  }\Big(\Phi_{0}+\frac{\Phi_{2}}{r^4}+\mathcal{O}\Big(\frac{1}{r^{6}}\Big)\Big)
\ee
where $\Phi_i (i=0,2\cdots)$s are some constant factors that appear in the expansion of the radial function $ F(r) $ near the boundary of the $ AdS_5 $. For the present purpose of our analysis it is sufficient to consider terms upto $\mathcal{O}(\frac{1}{r^4})$. 

  In order to compute the shear viscosity from the retarded Green's function we only need to register the leading fall off of $ F(r) $ corresponding to the radial coordinate ($ r $) at the leading order in $ \nu $ which finally yields,
\be
F(r)= \Phi_{0}+\frac{ \Phi_{2}}{ r^4}+ \frac{i \nu (1- 4\,\lambda) }{4 N_{\#}}\Big(\Phi_{0}+\frac{ \Phi_{2}}{ r^4}\Big) .\\
\ee
  In order to compute the retarded Green's function ($ G^{R}(\nu, \vec k=0) $) we only need to evaluate the following quantity namely \cite{Min},
\be
\mathcal{F}=\lim_{r \rightarrow \infty} \sqrt{g} g^{rr}\frac{F(r)}{\Phi_{0}\Phi_{2}f_{\infty}} \partial_{r} F(r)=-4N_{\#}-2 i\, \nu\, (1-4\,\lambda).\label{E20}
\ee

Using (\ref{E20}) we finally obtain the retarded Green's function as,
\be
G^{R}_{yz\,,\,yz}(\nu, \vec k=0)=\frac{1}{2\kappa_{5}^2}\mathcal{F}=-\frac{2N_{\#}}{\kappa_{5}^2} -\frac{i \nu (1-4\,\lambda)}{\kappa_{5}^2}.
\ee

Finally, the shear  viscosity $\eta_{yz}$ corresponding to the $y\ z$ plane turns out to be,
\be
\eta_{yz}=-\lim_{\nu \rightarrow 0}\frac{1}{2\nu } \operatorname{Im}(G^{R}_{yz\,,\,yz})=\frac{1-4\,\lambda}{2\kappa_{5}^2}.
\ee

Using (\ref{Anisotrophic entropy}), the shear viscosity to entropy ratio corresponding to the graviton fluctuations along $(y,z)$ plane turns out to be,
\be
\frac{\eta_{yz}}{s}= \frac{(1-4\,\lambda)}{4\pi}.
\ee

This is the famous $ \eta/s $ result for the EGB gravity known for quite a long time \cite{blmsy} and is quite expected from the physical arguments that we had mentioned earlier in Section 1. This result is due to the fact that the graviton modes along the $ (y,z) $ direction are essentially decoupled from the gauge degrees of freedom and therefore gives rise to the universal result as known for the EGB gravity. In the next section we are going to compute the $ \eta_{xy}/s$ ratio where due to the presence of the interaction between the graviton and the gauge degrees of freedom we expect some finite temperature corrections to the above universal result in EGB gravity.

\subsection{Calculation of $ \eta_{xy}/s$}
In order to compute the $ \eta_{xy}/s$ ratio we turn on graviton fluctuations $h_{xy} (= r^2 f^2(r)\Psi(r,t))$ as well as the gauge fluctuations namely $\delta A^{1}_{y}(r,t)$ and $\delta A^{2}_{y}(r,t)$.  Upper indices on $\delta A$ correspond to SU(2) indices and the lower indices correspond to space time index. Following the same procedure as before we Fourier transform all the variables as,
\begin{eqnarray}
\Psi(r,t)&=&\int^{\infty}_{-\infty}e^{-i\nu t}\Psi(r) d\nu\nonumber \\
\delta A^{1}_{y}(r,t)&=&\int^{\infty}_{-\infty}e^{-i\nu t}\delta A^{1}_{y}(r) d\nu \nonumber\\ 
\delta A^{2}_{y}(r,t)&=&\int^{\infty}_{-\infty}e^{-i\nu t}\delta A^{2}_{y}(r) d\nu.\label{E29}
\end{eqnarray}

Substituting (\ref{E29}) in to (\ref{E4}) we arrive at the following set of equations namely\footnote{See Appendix A for details.}, 
\begin{eqnarray}
\label{Psi-equation}
0&=&X(r)\Psi^{\prime\prime}(r)+Y(r)\Psi^{\prime}(r) +\Big(Z(r)+\nu^2 W(r)\Big)\Psi(r) \nonumber\\&+& 4 \alpha ^2 \omega '(r)\delta A_{y}^{1\prime}(r) -\frac{4 \alpha ^2 \omega (r) \phi (r)^2 \delta A_{y}^{1}(r)}{N_{\#}^2 N(r)^2 \sigma (r)^2} -\frac{4 i \alpha ^2 \nu  \omega (r) \phi (r)\delta A_{y}^{2}(r)}{N_{\#}^2 N(r)^2 \sigma (r)^2} \nonumber\\
\label{a1-equation}
0&=&\delta A_{y}^{1\prime\prime}(r)+\left( \frac{1}{r} -\frac{2f^{\prime}(r)}{f(r)} +\frac{N^{\prime}(r)}{N(r)} 
+ \frac{\sigma^{\prime}(r)}{\sigma(r)} \right)\delta A_{y}^{1\prime}(r)
+ \left( \frac{\nu^2-\phi^{2}(r)}{N_{\#}^2 N(r)^2\sigma^{2}(r)}  \right)\delta A_{y}^{1}(r)\nonumber\\
&-&\tilde W(r) \Psi(r)\nonumber\\
\label{a2-equation}
0&=&\delta A_{y}^{2\prime\prime}(r)+\left( \frac{2 f'(r)}{f(r)}+\frac{N'(r)}{N(r)}+\frac{\sigma '(r)}{\sigma (r)}+\frac{5}{r} \right)\delta A_{y}^{2\prime}(r) 
+\left( \frac{\nu ^2-\phi (r)^2}{N_{\#}^2 N(r)^2 \sigma (r)^2} \right)\delta A_{y}^{2}(r)\nonumber\\
&+&\frac{f(r)^4 \omega (r)^2}{r^2 N(r)}\delta A_{y}^{2}(r)  -\frac{i \nu  f(r)^4 \omega (r) \phi (r)}{2 N_{\#}^2 r^2 N(r)^2 \sigma (r)^2} \Psi(r). 
\end{eqnarray}

Considering the in-going boundary condition we take solutions of the following form namely,
\begin{eqnarray} \label{eq31}
\Psi(r)= &\left( \frac{N(r)}{r^2} \right)^{-i\frac{\nu \tilde T}{4}} F(r)\nonumber\\
\delta A^{1}_{y}(r)= &\left( \frac{N(r)}{r^2} \right)^{-i\frac{\nu \tilde T}{4}} H(r)\nonumber\\
\delta A^{2}_{y}(r)= &\left( \frac{N(r)}{r^2} \right)^{-i\frac{\nu \tilde T}{4}} J(r)
\end{eqnarray}
where $\tilde T$ is given in (\ref{tt}).\\
In the next step we substitute (\ref{eq31}) in to (\ref{a2-equation}) which yields \footnote{For details see Appendix B.},
\begin{eqnarray} \label{eq-32}
0&=&-\tilde U_{1}(r) F''(r)+\Big(\tilde X(r) +\frac{i \nu\, \tilde T\, \tilde Y(r)}{4}\Big)F'(r)+\Big (\tilde U(r)+\frac{i\,\nu\,\tilde T}{4}\tilde Z(r)-\frac{\nu^2 \,\tilde T^2}{16}\tilde V(r) \Big)F(r)\nonumber\\&-&\frac{\nu^2}{N_{\#}^2}\tilde X_{1}(r) F(r)+\alpha^2\Big(\frac{f(r)^4\omega'(r)^2}{6 r^2}-\frac{\phi'(r)^2}{6\,N_{\#}^2N(r)\sigma(r)^2}-\frac{f(r)^4\phi(r)^2\omega(r)^2}{6\,N_{\#}^2r^2 N(r)^2\sigma(r)^2}\Big)F(r)\nonumber\\&+&\frac{i\, \tilde T\,\nu\, \phi(r)\,\omega(r)}{N_{\#}^2N(r)^2\sigma(r)^2}\delta A^{2}_{y}(r)+\frac{\alpha^2\,\phi(r)^2\,\omega(r)}{N_{\#}^2 N(r)^2 \sigma(r)^2}\delta A^{1}_{y}(r)
\end{eqnarray}
\begin{eqnarray}
0&=&\delta A^{1''}_{y}(r)+\Big(\frac{N'(r)}{N(r)}+\frac{1}{r}+\frac{\sigma'(r)}{\sigma(r)}-\frac{2 f'(r)}{f(r)}\Big)\delta A^{1'}_{y}(r)-\frac{\phi(r)^2}{N_{\#}^2 N(r)^2\sigma(r)^2}\delta A^{1}_{y}(r)\nonumber\\&+&\frac{\nu^2}{N_{\#}^2 N(r)^2 \sigma(r)^2}\delta A^{1}_{y}(r)+\frac{i \,\nu\,\tilde T}{4} \Big(\frac{4}{r}-\frac{2 N'(r)}{N(r)}\Big) \delta A^{1'}_{y}(r)-\frac{\nu^2 \tilde T^2}{16}\Big(\frac{N'(r)^2}{N(r)^2}-\frac{ 4 N'(r)}{ r N(r)}+\frac{4}{r^2}\Big)\delta A^{1}_{y}(r)\nonumber\\&+&\tilde Y_{1}(r) F(r)-\frac{f(r)^4 \omega'(r)}{2 r^2} F'(r)+\frac{i\,\nu\,\tilde T}{4}\tilde Z_{1}(r)\delta A^{1}_{y}(r)+\frac{i\,\nu\,\tilde T}{4}\frac{f(r)^4 N'(r)\omega'(r)}{2\,r^2\,N(r)}F(r)
\end{eqnarray}
\begin{eqnarray}
0&=&\delta A^{2''}_{y}(r)+\frac{i\,\nu\,\tilde T}{4}\Big(\frac{4}{r}-\frac{2 N'(r)}{N(r)}\Big)\delta A^{2'}_{y}(r)+\frac{i\,\nu\,\tilde T}{4}\tilde W_{1}(r)\delta A^{2}_{y}(r)\nonumber\\&-&\frac{\nu^2\,\tilde T^2}{16}\Big(\frac{4}{r^2}-\frac{4\,N'(r)}{r\,N(r)}+\frac{N'(r)^2}{N(r)^2}\Big)\delta A^{2}_{y}(r)+\frac{\nu^2}{N_{\#}^2N(r)^2\sigma(r)^2}\delta A^{2}_{y}(r)-\frac{i\,\nu\,f(r)^4\phi(r)\omega(r)}{2\,N_{\#}^2\,r^2N(r)^2\sigma(r)^2}F(r)\nonumber\\&+&\Big(\frac{2\,f'(r)}{f(r)}+\frac{N'(r)}{N(r)}+\frac{\sigma'(r)}{\sigma(r)}+\frac{5}{r}\Big)\delta A^{2'}_{y}(r)+\Big(\frac{f(r)^4\omega(r)^2}{r^2\,N(r)}-\frac{\phi(r)^2}{N_{\#}^2N(r)^2\sigma(r)^2}\Big)\delta A^{2}_{y}(r).
\end{eqnarray}

Following the same procedure to find the retarded Green's function we expand $F(r), H(r)$ and $ J(r)$ upto linear order in the parameter $\nu$ namely, 
\begin{eqnarray}
F(r)&=& F_{0} (r)+\frac{i \nu}{4} F_{1}(r)+\mathcal{O}(\nu^2)\nonumber\\
H(r)&=& H_{0}(r)+\frac{i \nu}{4}H_{1}(r)+\mathcal{O}(\nu^2)\nonumber\\
J(r)&=&J_{0}(r)+\frac{i\nu}{4}J_{1}(r)+\mathcal{O}(\nu^2).\label{E33}
\end{eqnarray}

 Using the above expansion (\ref{E33}) we finally calculate the shear viscosity ($\eta_{xy}$) associated with the ($x,y$) plane. In order to do that we first substitute the expansion (\ref{E33}) in to (\ref{eq-32}) and note the overall fall off of the radial function $ F(r) $ at large values of the radial coordinate ($ r $).  As we have seen earlier in Section (3.1), in order to compute the shear viscosity ($ \eta_{xy} $) from the retarded Green's function we only need to register the fall off $ \mathcal{O}(1/r^{4}) $ in the radial coordinate ($ r $) at the leading order in $ \nu $ which finally yields,
\be
F(r)=  \Psi_{0}+\frac{\Psi_2}{r^4}+ \frac{i \nu\,(1-4\,\lambda) }{ 4 N_{\#} }\Big( 1+\varepsilon^2 \alpha^2 \frac{29 }{896 } \Big)\Big(\Psi_{0}+\frac{\Psi_2}{r^4}\Big)\label{E34} 
\ee
where $ \Psi_{i} $s are some constants as we have seen before.

At this stage it is indeed quite interesting to note that in the $\lambda=0$ limit, the above expression (\ref{E34}) reduces  correctly to its corresponding Einstein counterpart as given in \cite{Basu2}. 
Finally, using (\ref{E20}) we arrive at the expression for the shear viscosity to entropy ratio in the $(x,y)$ plane corresponding to the symmetry broken phase namely,
\be \label{T3}
 \frac{\eta_{xy}}{s}= \frac{1- 4\,\lambda}{4\pi}\Big(1 + \varepsilon^2\,\alpha^2 \frac{29}{896}\Big).
\ee

 Before we proceed further a few comments are in order. First of all, the presence of the $\varepsilon^2 \alpha^2$ term in the above expression (\ref{T3}) essentially corresponds to the fact that we are explicitly sitting in the symmetry broken phase of the system. One can easily identify this particular piece in the theory appearing due to some finite temperature corrections below certain critical temperature ($ T_c $).  As a matter of fact it is in fact quite evident from (\ref{T1}) and (\ref{E215}) that this $\varepsilon^2\alpha^2$ term is proportional to  $(T-T_{c})$ with certain proportionality factor $K(\lambda)$. Therefore we multiply both side of (\ref{T3}) by 
$ \frac{4\pi}{1-4\,\lambda}$ in order to extract out the $\varepsilon^2 \alpha^2$ factor which finally yields,
\be
\frac{4 \pi}{1-4\,\lambda} \frac{\eta_{xy}}{s}= 1 +K (\lambda)\,\,T_{c}\,\Big (1-\frac{T}{T_{c}}\Big)^{\beta}\label{E35}
\ee
where $\beta=1$ and, 
\be \label{eq-K}
K(\lambda)=-\frac{682515 \,\pi\,N_{\#} (3-16\, \lambda )}{32(32448-35N_{\#}^2(165-8368\,\lambda))}.
\ee
 Note that $\beta$ has to be equal to one as long as we stick to the order $\varepsilon^2\alpha^2$ in our calculations. 
 
At this stage it is noteworthy to mention that in the $\lambda=0$ limit Eq.(\ref{E35}) reduces to its corresponding Einstein counterpart as given in \cite{Basu2}. Speaking more specifically, Eq.(\ref{E35}) is the generalization of the earlier results \cite{Basu2} in the presence of higher derivative corrections. The point that we want to stress at this stage is the following: Like in the pure Einstein case, we note that the $ \eta/s $ ratio in particular in the EGB gravity is also not universal rather it explicitly contains the finite temperature corrections below certain critical temperature ($ T<T_c $). This is the first non trivial observation regarding the non universality of the $ \eta/s $ ratio in higher derivative theories of gravity to the best of our knowledge.

\section[Causality constraint]{Causality constraint \footnote{We thank Aninda Sinha for suggesting this.}}
In this section we will try to explore the causality constraints on the Gauss-Bonnet (GB) coupling both in the broken as well as the unbroken phase in presence of $SU(2)$ Yang-Mills matter.  To start with we investigate the unbroken phase where the boundary $ SO(3)$ symmetry is preserved.  In order to do that we turn on perturbations along $(y,z)$ plane and from the linearized equation of motion we calculate the group velocity of the graviton wave packets near the boundary of the $ AdS_{5} $ in the large momentum limit. Demanding that the group velocity of graviton wave packet must be less than the speed of light we get the constraint on Gauss-Bonnet (GB) coupling ($ \lambda $) following \cite{blmsy}.  We repeat the same analysis for the symmetry broken phase where besides the graviton fluctuations gauge field perturbations are also turned on.  

\subsection{Unbroken phase }
In this case we have only $A^{3}_{t}$ component of the gauge field turned on. In order to calculate the bound on $ \lambda $, we turn on graviton fluctuations along ($ y,z $) plane and retain terms upto $\alpha^2$ order. We take the graviton fluctuations of the following form namely,
\be  \label{eq37}
h_{yz}(r,t,x)= e^{-i\nu t+ i k_{1} x + ik_{r} r}.
\ee 

For the metric we take the following ansatz namely,
\begin{eqnarray}
N(r)&=&\Big[\frac{r^2}{2\lambda}\Big(1- \sqrt{1-4\, \lambda  \left(1-\frac{1}{r^4}\right)}\Big)-\alpha^2 \Big (\frac{32 \left(r^2-1\right)}{3 r^2 N_{\#}^2\sqrt{4 \lambda +(1-4 \lambda ) r^4}}\Big)\Big]\,,\nonumber\\ \omega(r)&=&0\,,~~ \sigma(r)=1\,,~~ f(r)=1\,.\\ \nonumber
\end{eqnarray}

 Substituting (\ref{eq37}) in to (\ref{E4}) we obtain the linearized  graviton equation of motion in large momentum ($k^{\mu}$) limit as, 
\be
-\frac{1}{N_{\#}^2 N(r)}\nu^2+ N(r)k_{r}^2+ \frac{\left(1-\lambda\, N''(r)\right)}{ r \left(r-\lambda\, N'(r)\right)} k_{3}^2=0.\label{E39}
\ee

Eq.(\ref{E39}) could be expressed as,
\be
g^{\mu\nu}k_{\mu}k_{\nu}=0
\ee
with the metric $g_{\mu\nu}$ of the following form namely,
\be
g_{\mu\nu} dx^{\mu}dx^{\nu}= N(r)N_{\#}^2\Big(-dt^2+\frac{1}{c_{g}^2} dx^2\Big)+\frac{1}{N(r)} dr^2
\ee
where,
\be
c_{g}^2= \frac{N_{\#}^2 N(r)}{r^2} \frac{1-\lambda\, N''(r)}{1-\frac{\lambda\, N'(r)}{r}}
\ee
is the group velocity of the graviton wave packet.

To avoid any acausal propagation at the boundary we should have $ c_{g}^2 \leq 1 $ which finally leads to the bound on the GB coupling ($ \lambda $) namely,
\be
\lambda \leq \frac{9}{100}\,.\\
\ee
This matches with the earlier results of \cite{blmsy, hm} in the absence of any matter couplings.  In this case  $\omega=0$ i.e only the temporal component of the gauge is field is turned one. So effectively we have a $U(1)$ gauge field coupled with gravity.  Our analysis shows that in this case also we get the same bound as what one gets in absence of any matter fields. This translates into the following well known  statement of $\eta_{yz}/s $ namely,
\be
\frac{\eta_{yz}}{s}\geq \frac{4}{25\pi}\,.\\
\ee

\subsection{Broken phase }
In order to study the bound on GB coupling ($ \lambda $) corresponding to the symmetry broken phase we turn on graviton fluctuations $(h_{xy}(r,t,z))$ as well as the gauge fluctuations namely $\delta A^{1}_{y}(r,t,z)$ and $\delta A^{2}_{y}(r,t,z)$.  Under these circumstances the graviton fluctuations will mix with the gauge field perturbations and as a result we have to consider the full back reacted metric as mentioned in (\ref{Sol1}).  We take the graviton fluctuations of the following form namely,
\be \label{eq38}
h_{xy}(r,t,z)= e^{-i\nu t+ i k_{3} z + ik_{r} r}.\\
\ee 

Also we take  the gauge field perturbations of the form,
\be \label{eq39}
\delta A^{2}_{y}(r,t,z)=\delta A^{1}_{y}(r,t,z)=e^{-i\tilde\nu t+ i \tilde k_{3} z + i\tilde k_{r} r}.\\
\ee
   Using the profile mentioned in (\ref{eq38}), the linearized  graviton equation of motion in large momentum ($k^{\mu}$) limit turns out to be, 
\be
-\frac{1}{N(r) N_{\#}^2\sigma(r)^2} \nu^2+g^{rr}  k_{r}^2+g^{xx} k_{3}^2=0\,.
\ee

This could be further rewritten as,
\be
g^{\mu\nu}k_{\mu}k_{\nu}=0\\
\ee
where the metric $g_{\mu\nu}$ could be expressed as,
\be
g_{\mu\nu} dx^{\mu}dx^{\nu}= N(r) N_{\#}^2\sigma(r)^2\Big(-dt^2+\frac{1}{c_{g}^2} dx^2\Big)+g_{rr} dr^2
\ee
with,
\begin{eqnarray}
g_{rr}&=&\frac{\sigma (r) \left(\lambda\,  \left(r f'(r) N'(r)+2 N(r) \left(r f''(r)+2 f'(r)\right)\right)+f(r) \left(\lambda\, N'(r)-r\right)\right)}{N(r) \left(\lambda \, r f'(r) \left(\sigma (r) N'(r)+2 N(r) \sigma '(r)\right)+f(r) \left(\sigma (r) \left(\lambda\,  N'(r)-r\right)+2 \lambda\,  N(r) \sigma '(r)\right)\right)}\nonumber\\
c_{g}^2&=&\frac{N_{\#}^2 N(r) \sigma (r) \left(-\lambda\,  \sigma (r) N''(r)-3 \lambda N'(r) \sigma '(r)-2 \lambda\, N(r) \sigma ''(r)+\sigma (r)\right)}{r f(r) \left(-2 \lambda \, r N(r) f''(r)-\lambda\,  f'(r) \left(r N'(r)+4 N(r)\right)+f(r) \left(r-\lambda \, N'(r)\right)\right)}.
\end{eqnarray}

Finally expanding $ c_g^{2} $ around the boundary $r\rightarrow \infty$ and demanding the fact that the leading term has to be less than or equal to zero we note that the GB coupling ($\lambda$) has the following upper bound even in the symmetry broken phase namely,
\be
\lambda \leq \frac{9}{100}.\label{lambda}
\ee

This result is indeed quite surprising. Although in the symmetry broken phase we have the full $SU(2)$ sector of the gauge field turned on, still it is quite surprising to note that one gets the same upper bound on the GB coupling ($ \lambda $) as we found earlier corresponding to the symmetry unbroken phase. Note that the above result is valid only in certain special circumstances in the sense that on one hand we have treated the GB coupling ($\lambda$) non perturbatively and on the other hand we have treated the $ SU(2) $ gauge sector perturbatively near the critical point ($ T\sim T_c $).  This shows that at least in this perturbative framework the bound on $\lambda$ doesn't change even in the presence of a non-abelian gauge coupling. Mathematically it can be justified as follows, if one looks at the linearized Gauss-Bonnet equation of motion (\ref{Psi-equation}) carefully, even though there is in general a mixing between the graviton and gauge fluctuations but in the high frequency limit they seem to get decoupled. It is evident from the linearized Gauss-Bonnet equation  (\ref{Psi-equation}) that $h_{xy}(r,t,z)$ comes with double derivative with respect to $r$, $t$ and $z$ but the gauge fluctuations $\delta A^{1}_{y}(r,t,z)$ and $\delta A^2_{y}(r,t,z)$ come with no derivatives. So naturally in the high frequency limit terms with double derivatives dominate over those terms with lesser number of derivatives. As a consequence of this the graviton fluctuations decouple from the gauge fluctuations. Hence we get the same bound although there are gauge fields present. At this stage it is worthwhile to mention that one might get some nontrivial result in the case when both the gauge and gravitational sectors are treated non perturbatively. We leave this issue as a part of future investigations.

The key observation from the above analysis is that the proportionality factor $K(\lambda)$ (in equation (\ref{eq-K})) corresponding to the above (upper) bound (\ref{lambda}) of GB coupling ($ \lambda $) turns out to be,
\be
K(\lambda)=-\sqrt{\frac{41}{2}}\frac{682515\, \pi}{5059184} \approx -1.91893~~.
\ee
As a result the corresponding lower bound for $\eta_{xy}/s$ changes to,
\be
\frac{\eta_{xy}}{s}\geq\frac{4}{25\,\pi}-0.09773\,\, T_{c}\,\Big(1-\frac{T}{T_{c}}\Big)^{\beta}
\ee
with $\beta=1\,.$

\section{Summary and final remarks}

Before we conclude this article, it is now a good time to summarize all the crucial results/findings of the present analysis.
 One of the major outcomes of the present analysis is the finding of the non universal shear viscosity to entropy ($ \eta/s $) ratio (corresponding to the super fluid phase) in higher derivative gravity. This deviation is caused due to the mutual interactions of the helicity two modes to that with the helicity one modes in the symmetry broken phase.
 
 Another important outcome of our analysis is the fact that the upper bound of the Gauss- Bonnet coupling ($ \lambda $) does not seem to get changed even in the super fluid phase as long as we are close to the critical point ($ T\sim T_c $) of the phase diagram. However we note that along the symmetry broken direction the corresponding lower bound of $ \eta/s $ ratio gets modified due to the presence of finite temperature corrections. It would be an interesting exercise to explore by what amount this upper bound on $ \lambda $ is changed as we move away from the critical point or we treat the $ p $- wave order parameter non perturbatively. Also it will be interesting and insightful to repeat this analysis for other anisotropic models, for example one can look at the models mentioned in \cite{Mateos}.

\section*{\bf {Acknowledgements }}
We are grateful to Aninda Sinha for valuable suggestions and remarks and for going through the manuscript in details. DR would like to acknowledge the financial support from CHEP, Indian Institute of Science, Bangalore.\\

\appendix

\section{Details of linearized equations of motion for symmetry unbroken phase}

In this appendix we give the detailed expressions for $X(r),  Y(r),  Z(r),  W(r)$ and $\tilde W(r)$ as mentioned in equation (\ref{Psi-equation}).
\begin{eqnarray}
X(r)&=&1-\frac{\lambda  f'(r) N'(r)}{f(r)}-\frac{2 \lambda \, N(r) f'(r) \sigma '(r)}{f(r) \sigma (r)}-\frac{\lambda  N'(r)}{r}-\frac{2 \lambda  N(r) \sigma '(r)}{r \sigma (r)}\,,\nonumber\\
W(r)&=&\frac{1}{N_{\#}^2 N(r)^2 \sigma (r)^2}-\frac{2 \lambda  f''(r)}{N_{\#}^2 f(r) N(r) \sigma (r)^2}-\frac{\lambda  f'(r) N'(r)}{N_{\#}^2 f(r)N(r)^2 \sigma (r)^2}\nonumber\\&-&\frac{4 \lambda  f'(r)}{N_{\#}^2 r f(r) N(r) \sigma (r)^2}- \frac{\lambda N'(r)}{N_{\#}^2 r N(r)^2 \sigma (r)^2}\,,\nonumber\\
\tilde W(r)&=&\frac{f(r)^3 f'(r) \omega '(r)}{r^2}+\frac{f(r)^4 N'(r) \omega '(r)}{2 r^2 N(r)} -\frac{f(r)^4 \omega (r) \phi (r)^2}{2 N_{\#}^2 r^2 N(r)^2 \sigma (r)^2}-\frac{f(r)^4 \omega '(r)}{2 r^3} \nonumber\\ &+&\frac{f(r)^4 \sigma '(r) \omega '(r)}{2 r^2 \sigma (r)}+\frac{f(r)^4 \omega ''(r)}{2 r^2}\,.
\end{eqnarray}
\begin{eqnarray}
Y(r)&=& \frac{\sigma '(r)}{\sigma (r)}+\frac{2 f'(r)}{f(r)}+\frac{N'(r)}{N(r)}-\frac{1}{r}-\frac{\lambda  f''(r) N'(r)}{f(r)}-\frac{2 \lambda N(r) f''(r) \sigma '(r)}{f(r) \sigma (r)}\nonumber\\&-&\frac{\lambda  f'(r) N''(r)}{f(r)}-\frac{5 \lambda  f'(r) N'(r) \sigma '(r)}{f(r) \sigma (r)}-\frac{\lambda  f'(r)^2 N'(r)}{f(r)^2}-\frac{\lambda  f'(r) N'(r)}{r f(r)}\nonumber\\&-&\frac{\lambda  f'(r) N'(r)^2}{f(r) N(r)}-\frac{2 \lambda  N(r) f'(r) \sigma ''(r)}{f(r) \sigma (r)}- \frac{2 \lambda N(r) f'(r)^2 \sigma '(r)}{f(r)^2 \sigma (r)}-\frac{2 \lambda N(r) f'(r) \sigma '(r)}{r f(r) \sigma (r)}\nonumber\\&-&\frac{\lambda  N''(r)}{r}+\frac{2 \lambda  N'(r)}{r^2}-\frac{5 \lambda  N'(r) \sigma '(r)}{r \sigma (r)}-\frac{\lambda N'(r)^2}{r N(r)}+\frac{4 \lambda N(r) \sigma '(r)}{r^2 \sigma (r)}-\frac{2 \lambda N(r) \sigma ''(r)}{r \sigma (r)}\,,\nonumber\\ 
Z(r)&=&\frac{4}{r^2}-\frac{8}{N(r)}-\frac{4 f'(r)}{r f(r)}-\frac{8 f'(r)^2}{f(r)^2}+\frac{2 \alpha ^2 f(r)^4 \omega (r)^2 \phi (r)^2}{3 N_{\#}^2 r^2 N(r)^2 \sigma (r)^2}-\frac{2 \alpha ^2 f(r)^4 \omega '(r)^2}{3 r^2}\nonumber\\&+&\frac{2 \alpha ^2 \phi '(r)^2}{3 N_{\#}^2 N(r) \sigma (r)^2}+\frac{2 \lambda  f''(r) N'(r)}{r f(r)}+\frac{4 \lambda N(r) f''(r) \sigma '(r)}{r f(r) \sigma (r)}+\frac{2 \lambda  f'(r) N''(r)}{r f(r)}\nonumber \\&+&\frac{4 \lambda  f'(r) N'(r)}{r^2 f(r)}+\frac{10 \lambda  f'(r) N'(r) \sigma '(r)}{r f(r) \sigma (r)}+\frac{2 \lambda  f'(r) N'(r)^2}{r f(r) N(r)}+\frac{16 \lambda  f'(r)^3 N'(r)}{f(r)^3}\nonumber\\&+&\frac{22 \lambda  f'(r)^2 N'(r)}{r f(r)^2}+\frac{8 \lambda N(r) f'(r) \sigma '(r)}{r^2 f(r) \sigma (r)}+\frac{8 \lambda N(r) f'(r)^2}{r^2 f(r)^2}+\frac{4 \lambda  N(r) f'(r) \sigma ''(r)}{r f(r) \sigma (r)}\nonumber\\&+&\frac{24 \lambda  N(r) f'(r)^3 \sigma '(r)}{f(r)^3 \sigma (r)}+\frac{32 \lambda N(r) f'(r)^2 \sigma '(r)}{r f(r)^2 \sigma (r)}-\frac{8 \lambda N(r) f'(r)^4}{f(r)^4}+\frac{8 \lambda N(r) f'(r)^2 f''(r)}{f(r)^3}\nonumber\\&+&\frac{8 \lambda  N(r) f'(r) f''(r)}{r f(r)^2}+\frac{2 \lambda N''(r)}{r^2}-\frac{6 \lambda  N'(r)}{r^3}+\frac{10 \lambda N'(r) \sigma '(r)}{r^2 \sigma (r)}+\frac{2 \lambda N'(r)^2}{r^2 N(r)}-\nonumber\\&&\frac{8 \lambda N(r) \sigma '(r)}{r^3 \sigma (r)}+\frac{4 \lambda N(r) \sigma ''(r)}{r^2 \sigma (r)}\,,
\end{eqnarray}

\section{Details of linearized equations of motion for symmetry broken phase}

In this appendix we give the detailed expressions for $\tilde X(r)$, $\tilde Y(r)$, $\tilde Z(r)$, $\tilde U(r)$, $\tilde V(r)$, $\tilde X_{1}(r)$,\\$\tilde Y_{1}(r), \tilde U_{1}(r)$, $ \tilde Z_{1}(r)$ and $\tilde W_{1}(r)$ as mentioned in equation (\ref{eq-32}).

\begin{eqnarray}
\tilde Y(r)&=&-\frac{1}{r}+\frac{N'(r)}{N(r)}+\frac{\lambda\,N'(r)}{r^2}+\frac{\lambda\,f'(r)\,N'(r)}{r\,f(r)}-\frac{\lambda\, N'(r)^2}{2\,r\,N(r)}-\frac{\lambda\,f'(r)\,N'(r)^2}{2\,f(r)\,N(r)}+\frac{2\,\lambda\,N(r)\,\sigma'(r)}{r^2\,\sigma(r)}\nonumber\\&+&\frac{2\,\lambda\,N(r)\,f'(r)\,\sigma'(r)}{r\,f(r)\,\sigma(r)}-\frac{\lambda\,N'(r)\sigma'(r)}{r\,\sigma(r)}-\frac{\lambda\,f'(r)\,N'(r)\,\sigma'(r)}{f(r)\,\sigma(r)}\,.
\end{eqnarray}
\begin{eqnarray}
\tilde X(r)&=&\frac{1}{4\,r}-\frac{f'(r)}{2\,f(r)}-\frac{N'(r)}{4\,N(r)}-\frac{\sigma'(r)}{4\,\sigma}-\frac{\lambda\,N'(r)}{2\,r^2}+\frac{\lambda\,f'(r)\,N'(r)}{4\,r\,f(r)}+\frac{\lambda\,f'(r)^2N'(r)}{4\,f(r)^2}\nonumber\\&+&\frac{\lambda\,N'(r)^2}{4\,r\,N(r)}+\frac{\lambda\,f'(r)\,N'(r)^2}{4\,f(r)\,N(r)}-\frac{\lambda\,N(r)\,\sigma'(r)}{r^2\,\sigma(r)}+\frac{\lambda\,N(r)f'(r)\sigma'(r)}{2\,r\,f(r)\sigma(r)}+\frac{\lambda\,N(r)\,f'(r)^2\sigma'(r)}{2\,f(r)^2\,\sigma(r)}\nonumber\\&+&\frac{5\,\lambda\,N'(r)\sigma'(r)}{4\,r\,\sigma(r)}+\frac{5\,\lambda\,f'(r)N'(r)\sigma'(r)}{4\,f(r)\,\sigma(r)}+\frac{\lambda\,N'(r)\,f''(r)}{4\,f(r)}+\frac{\lambda\,N(r)\,\sigma'(r)f''(r)}{2\,f(r)\sigma(r)}+\frac{\lambda\,N''(r)}{4\,r}\nonumber\\&+&\frac{\lambda\,f'(r)\,N''(r)}{4\,f(r)}+\frac{\lambda\,N(r)\,\sigma''(r)}{2\,r\,\sigma(r)}+\frac{\lambda\,N(r)\,f'(r)\,\sigma''(r)}{2\,f(r)\,\sigma(r)}\,,\nonumber\\
\end{eqnarray}
\begin{eqnarray}
\tilde V(r)&=&-\frac{1}{r^2}+\frac{N'(r)}{N(r) r}-\frac{N'(r)^2}{4\,N(r)^2}+\frac{\lambda\,N'(r)}{r^3}+\frac{\lambda\,f'(r)\,N'(r)}{r^2\,f(r)}-\frac{\lambda\,N'(r)^2}{r^2\,N(r)}-\frac{\lambda\,f'(r)N'(r)^2}{r\,N(r)\,f(r)}\nonumber\\&+&\frac{\lambda\,N'(r)^3}{4\,r\,N(r)^2}+\frac{\lambda\,f'(r)\,N'(r)^3}{4\,f(r)\,N(r)^2}+\frac{2\,\lambda\,N(r)\,\sigma'(r)}{r^3\,\sigma(r)}+\frac{2\,\lambda\,N(r)\,f'(r)^2\,\sigma'(r)^2}{r^2\,f(r)\,\sigma(r)}-\frac{2\lambda\,N'(r)\sigma'(r)}{r^2\sigma(r)}\nonumber\\&-&\frac{2\lambda\,f'(r)\,N'(r)\,\sigma(r)}{r\,f(r)\,\sigma(r)}+\frac{\lambda\,N'(r)^2\,\sigma'(r)}{2\,r\,N(r)\sigma(r)}+\frac{\lambda\,f'(r)\,N'(r)^2\,\sigma'(r)}{2\,f(r)\,\sigma(r)}\,.
\end{eqnarray}
\begin{eqnarray}
\tilde Z(r)&=&\frac{1}{r^2}-\frac{f'(r)}{r\,f(r)}-\frac{3\,N'(r)}{4\,r\,N(r)}+\frac{f'(r)\,N'(r)}{2\,N(r)\,f(r)}-\frac{\sigma'(r)}{2\,r\,N(r)\,\sigma(r)}+\frac{N'(r)\sigma'(r)}{4\,N(r)\,\sigma(r)}+\frac{N''(r)}{4\,N(r)}\nonumber\\&-&\frac{3 N'(r)}{4\,r\,N(r)}-\frac{3\,\lambda\,N'(r)}{2\,r^3}+\frac{\lambda\,f'(r)^2 N'(r)}{2\,r\,f(r)^2}+\frac{\lambda N'(r)^2}{r^2\, N(r)}+\frac{\lambda\,f'(r)\,N'(r)^2}{4\,r\,N(r)\,f(r)}-\frac{\lambda\,f'(r)^2\,N'(r)^2}{4\,N(r)\,f(r)^2}\nonumber\\&-&\frac{3\,\lambda\,N(r)\,\sigma'(r)}{r^3\,\sigma(r)}+\frac{\lambda\,N(r)\,f'(r)^2\,\sigma'(r)}{r\,f'(r)^2\,\sigma(r)}+\frac{7\,\lambda\,N'(r)\,\sigma'(r)}{2\,r^2\,\sigma(r)}+\frac{2\,\lambda\,f'(r)\,N'(r)\,\sigma'(r)}{r\,f(r)\,\sigma(r)}\nonumber\\&-&\frac{\lambda\,f'(r)^2N'(r)\sigma'(r)}{2\,f(r)^2\,\sigma(r)}-\frac{3\,\lambda\,N'(r)^2\,\sigma'(r)}{4\,r\,N(r)\,\sigma(r)}-\frac{3\lambda\,f'(r)N'(r)^2\,\sigma'(r)}{4\,N(r)\,f(r)\,\sigma(r)}+\frac{\lambda\,N'(r)\,f''(r)}{2\,r\,f(r)}\nonumber\\&-&\frac{\lambda N'(r)^2 f''(r)}{2\,r\,N(r)\,f(r)}-\frac{\lambda\,f'(r)\,N'(r)\,N''(r)}{2\,N(r)\,f(r)}+\frac{\lambda\,N(r)\,\sigma'(r)\,f''(r)}{r\,f(r)\,\sigma(r)}-\frac{\lambda\,N'(r)\sigma'(r) f''(r)}{2 f(r)\,\sigma(r)}\nonumber\\&+&\frac{\lambda\,N''(r)}{2\,N(r)\, r^2}+\frac{\lambda\, f'(r)\,N''(r)}{2\,r\,f(r)}-\frac{\lambda\, N'(r)\,N''(r)}{2\,r\,N(r)}-\frac{\lambda\,\sigma'(r)\,N''(r)}{2\,r\,\sigma(r)}-\frac{\lambda\,f'(r)\sigma'(r)\,N''(r)}{2\,f(r)\,\sigma(r)}\nonumber\\&+&\frac{\lambda\,N(r)\,\sigma''(r)}{r^2\,\sigma(r)}+\frac{\lambda\,N(r)\,f'(r)\,\sigma''(r)}{r\,f(r)\,\sigma(r)}-\frac{\lambda\,N'(r)\,\sigma''(r)}{2\,r\,\sigma(r)}-\frac{\lambda\, f'(r)\,N'(r)\,\sigma''(r)}{2\,f(r)\,\sigma(r)}\,.
\end{eqnarray}
\begin{eqnarray}
\tilde U(r)&=&\frac{2}{N(r)}-\frac{1}{r^2}+\frac{f'(r)}{r\,f(r)}+\frac{2\,f'(r)^2}{f(r)^2}-\frac{2\,\lambda\,N(r)\,f'(r)^2}{r^2\,f(r)^2}+\frac{2\,\lambda\,N(r) f'(r)^2}{f(r)^4}+\frac{3\,\lambda\,N'(r)}{2\,r^3}\nonumber\\&-&\frac{\lambda\, f'(r)N'(r)}{r^2\,f(r)}-\frac{11\,\lambda\,f'(r)^2\,N'(r)}{2\,r\,f(r)^2}-\frac{4\,\lambda\,f'(r)^3\,N'(r)}{f(r)^3}-\frac{\lambda\,N'(r)^2}{2\,r^2\,N(r)}-\frac{\lambda\,f'(r)\,N'(r)^2}{2\,r\,N(r)\,f(r)}\nonumber\\&+&\frac{2\,\lambda\,N(r)\,\sigma'(r)}{r^3\,\sigma(r)}-\frac{2\,\lambda\,N(r)\,f'(r)\,\sigma'(r)}{r^2\,f(r)\,\sigma(r)}-\frac{8\,\lambda\,N(r)\,f'(r)^2\,\sigma'(r)}{r\,f(r)^2\,\sigma(r)}-\frac{6\,\lambda N(r)f'(r)^3\sigma'(r)}{f(r)^3\,\sigma(r)}\nonumber\\&-&
\frac{5\,\lambda\,N'(r)\,\sigma'(r)}{2\,r^2\,\sigma(r)}-\frac{5\,\lambda\,f'(r)\,N'(r)\,\sigma'(r)}{2\,r\,f(r)\,\sigma(r)}-\frac{2\,\lambda\,N(r)\,f'(r)\,f''(r)}{r\,f(r)^2}-\frac{2\,\lambda\,N(r) f'(r)^2\,f''(r)}{f(r)^3}\nonumber\\&-&\frac{\lambda\, N'(r)\,f''(r)}{2\,r\,f(r)}-\frac{\lambda\,N(r)\,\sigma'(r)\,f''(r)}{r\,f(r)\,\sigma(r)}-\frac{\lambda\,N''(r)}{2\,r^2}-\frac{\lambda\,f'(r)\,N''(r)}{2\,r\,f(r)}-\frac{\lambda\,N(r)\,\sigma''(r)}{r^2\,\sigma(r)}\nonumber\\&-&\frac{\lambda N(r)\,f'(r)\,\sigma''(r)}{r\,f(r)\,\sigma(r)}\,.\nonumber\\
\end{eqnarray}
\begin{eqnarray}
\tilde X_{1}(r)&=&\frac{1}{4N(r)^2\sigma(r)^2}-\frac{\lambda\,f'(r)}{r\,N(r)\,f(r)\sigma(r)^2}-\frac{\lambda\,N'(r)}{4\,r\,N(r)^2\sigma(r)^2}-\frac{\lambda\, f'(r)N'(r)}{4\,f(r)N(r)^2\sigma(r)^2}-\nonumber\\&&\frac{\lambda\,f''(r)}{2\, N(r) f(r) \sigma(r)^2}\nonumber\,,\\
\end{eqnarray}
\begin{eqnarray}
\tilde Y_{1}(r)&=&\frac{f(r)^4 \phi(r)^2\omega(r)}{2N_{\#}^2 r^2 N(r)^2\sigma(r)^2}+\frac{f(r)^4 \omega'(r)}{2 r^3}-\frac{f(r)^3 f'(r)\omega'(r)}{r^2}-\frac{f(r)^4 N'(r)\omega'(r)}{2 r^2 N(r)}-\nonumber\\&& \frac{f(r)^4 \sigma'(r)\omega'(r)}{2 r^2\sigma(r)}-\frac{f(r)^4\omega''(r)}{2 r^2}\nonumber\,,\\
\tilde Z_{1}(r)&=&\frac{N'(r)}{r\,N(r)}-\frac{4\,f'(r)}{r\,f(r)}+\frac{2\, f'(r) N'(r)}{f(r) N(r)}+\frac{2\sigma'(r)}{r \,\sigma(r)}-\frac{N'(r)\sigma'(r)}{N(r)\sigma(r)}-\frac{N''(r)}{N(r)}\nonumber\,,\\
\tilde W_{1}(r)&=&\frac{8}{r^2}+\frac{4\,f'(r)}{r\,f(r)}-\frac{3\,N'(r)}{r\,N(r)}-\frac{2\,f'(r)\,N'(r)}{f(r)\,N(r)}+\frac{2 \,\sigma'(r)}{r\,\sigma(r)}-\frac{N'(r)\sigma'(r)}{N(r)\sigma(r)}-\frac{N''(r)}{N(r)}\nonumber\,,\\
\tilde U_{1}(r)&=&\frac{1}{4}-\frac{\lambda\,N'(r)}{4 r}-\frac{\lambda\, f'(r) N'(r)}{4 f(r)}-\frac{\lambda\,N(r) \sigma'(r)}{2 \,r\,\sigma(r)}-\frac{\lambda\,N(r) f'(r)\sigma'(r)}{2 f(r)\,\sigma(r)}
\end{eqnarray}

\end{document}